\documentclass[12pt]{article}
\usepackage{amssymb}
\usepackage{amsmath}
\usepackage[unicode,bookmarks,bookmarksopen,bookmarksopenlevel=2,colorlinks,linkcolor=blue,citecolor=green]{hyperref}

\input{tcilatex}
\begin{document}

\title{\textbf{Integrable Dispersive Chains and Energy Dependent Schr\"{o}%
dinger Operator}}
\author{Maxim V.~Pavlov$^{1,2}$ \\
$^{1}$Sector of Mathematical Physics,\\
Lebedev Physical Institute of Russian Academy of Sciences,\\
Leninskij Prospekt 53, 119991 Moscow, Russia\\
$^{2}$Laboratory of Geometric Methods in Mathematical Physics,\\
Lomonosov Moscow State University,\\
Leninskie Gory 1, 119991 Moscow, Russia }
\date{}
\maketitle

\begin{abstract}
In this paper we consider integrable dispersive chains associated with the
so called \textquotedblleft Energy Dependent\textquotedblright\ Schr\"{o}%
dinger operator. In a general case multi component reductions of these
dispersive chains are new integrable systems, which are characterised by two
arbitrary natural numbers. Also we show that integrable three dimensional
linearly degenerate quasilinear equations of a second order possess
infinitely many differential constraints. Corresponding dispersive
reductions are integrable systems associated with the \textquotedblleft
Energy Dependent\textquotedblright\ Schr\"{o}dinger operator.


\bigskip

\end{abstract}

\textit{on the occasion of the 77th birthday of Professor Alexey Borisovich
Shabat}

\tableofcontents

\section{Introduction}

The remarkable Korteweg de Vries equation is associated with the linear Schr%
\"{o}dinger equation%
\begin{equation*}
\psi _{xx}=(\lambda +u)\psi .
\end{equation*}%
This paper is devoted to description of integrable systems associated with
more general linear equation $\psi _{xx}=u\psi $, where a dependence $%
u(x,t,\lambda )$ with respect to the spectral parameter $\lambda $ can be
much more complicated (for instance, rational). According to \cite{antford}
we call such a linear equation $\psi _{xx}=u(x,t,\lambda )\psi $ the energy
dependent Schr\"{o}dinger equation.

The function $\psi (x,t,\lambda )$ satisfies the pair of \textit{linear}
equations in partial derivatives%
\begin{equation}
\psi _{xx}=u\psi ,\text{ \ }\psi _{t}=a\psi _{x}-\frac{1}{2}a_{x}\psi .
\label{g}
\end{equation}%
Then the compatibility condition $(\psi _{xx})_{t}=(\psi _{t})_{xx}$ yields
the relationship%
\begin{equation}
u_{t}=\left( -\frac{1}{2}\partial _{x}^{3}+2u\partial _{x}+u_{x}\right) a
\label{main}
\end{equation}%
between functions $u(x,t,\lambda )$ and $a(x,t,\lambda )$.

If we choose the linear dependences $u(x,t,\lambda )=\lambda +u^{1}(x,t)$
and $a(x,t,\lambda )=\lambda +a_{1}(x,t)$, we obtain nothing but the famous
Korteweg de Vries equation\footnote{%
An accurate derivation leads to $a_{1}=-\frac{1}{2}u_{1}+\xi $, where $\xi $
is an arbitrary constant. However, corresponding equation $u_{1,t}=\frac{1}{4%
}u_{1,xxx}-\frac{3}{2}u_{1}u_{1,x}+\xi u_{1,x}$ reduces to the form (\ref%
{kdv}) under the transformation $t\rightarrow t,x+\xi t\rightarrow x$.}%
\begin{equation}
u_{t}^{1}=\frac{1}{4}u_{xxx}^{1}-\frac{3}{2}u^{1}u_{x}^{1},  \label{kdv}
\end{equation}%
where $a_{1}=-\frac{1}{2}u^{1}$.

If we choose the quadratic dependence\footnote{%
Here we follow S.J. Alber, see \cite{SAlber}} $u(x,t,\lambda )=\lambda
^{2}+\lambda u^{1}(x,t)+u^{2}(x,t)$ and again the linear dependence $%
a(x,t,\lambda )=\lambda +a_{1}(x,t)$, we obtain nothing but the well-known
Kaup--Boussinesq system%
\begin{equation*}
u_{t}^{1}=-u^{1}u_{x}^{1}+\left( -\frac{1}{2}u^{1}+\xi \right)
u_{x}^{1}+u_{x}^{2},\text{ \ }u_{t}^{2}=\frac{1}{4}%
u_{xxx}^{1}-u^{2}u_{x}^{1}+\left( -\frac{1}{2}u^{1}+\xi \right) u_{x}^{2},
\end{equation*}%
where $a_{1}=-\frac{1}{2}u^{1}+\xi $. The arbitrary constant $\xi $ can be
removed by a linear change of independent variables ($t\rightarrow t,x+\xi
t\rightarrow x$) exactly as in the previous case. Thus finally the
Kaup--Boussinesq system takes the form%
\begin{equation}
u_{t}^{1}=u_{x}^{2}-\frac{3}{2}u^{1}u_{x}^{1},\text{ \ }u_{t}^{2}=\frac{1}{4}%
u_{xxx}^{1}-u^{2}u_{x}^{1}-\frac{1}{2}u^{1}u_{x}^{2}.  \label{kb}
\end{equation}

Multi-component \textit{polynomial} (with respect to the spectral parameter $%
\lambda $) generalization $u(x,t,\lambda )=\lambda ^{M}+\lambda
^{M-1}u^{1}(x,t)+...+u^{M}(x,t)$ was investigated in several papers \cite%
{antford}. Multi-component \textit{rational} (with respect to the spectral
parameter $\lambda $) generalization ($\epsilon _{k}$ are arbitrary
parameters)%
\begin{equation}
u(x,t,\lambda )=\frac{\lambda ^{M}u^{0}(x,t)+\lambda
^{M-1}u^{1}(x,t)+...+u^{M}(x,t)}{\epsilon _{M}\lambda ^{M}+\epsilon
_{M-1}\lambda ^{M-1}+...+\epsilon _{0}}  \label{racio}
\end{equation}%
was studied in \cite{antford2}. The authors considered two main subclasses
selected by the conditions: $\epsilon _{M}=0$ and $u^{0}=1$ (the so called
\textquotedblleft Generalized KdV type systems\textquotedblright ); $%
\epsilon _{M}=0$ but $u^{M}=1$ (the so called \textquotedblleft Generalized
Harry Dym type systems\textquotedblright ). In another paper \cite{marvan}
we found a third narrow subclass determined by a sole restriction $u^{M}=0$.

This paper is devoted to an open question: description of multi-component
integrable systems associated with different dependencies $u(x,t,\lambda )$
with respect to the spectral parameter $\lambda $. In this paper we
construct infinitely many multi component dispersive reductions of
\textquotedblleft integrable dispersive chains\textquotedblright\ introduced
in \cite{ShabatJNMP}. Thus we found infinitely many new multi-component
integrable systems. They can be written in a compact form (\ref{disp}), (\ref%
{a1}) and (\ref{aksi}). Existence of dispersive reductions for linearly
degenerate integrable hydrodynamic chains was discovered in \cite{mas}. The
hierarchy of these hydrodynamic chains is associated with five integrable
three dimensional quasilinear equations of a second order (see, for
instance, \cite{fermos}). First such a three dimensional quasilinear
equation of a second order can be written in a hydrodynamic form (\ref{three}%
). This equation possesses $N$ component hydrodynamic reductions
parameterised by $N$ arbitrary functions of a single variable (see \cite%
{maksjmp}). In this paper we show that integrable three dimensional
quasilinear system of a first order (\ref{three}) also possesses $N$
component dispersive reductions, which are integrable dispersive systems (%
\ref{disp}), (\ref{diss}) characterised by two arbitrary natural numbers.

The structure of the paper is as follows. In Section~\ref{sec:dispersive} we
considered general properties of integrable dispersive chains (associated
with the energy dependent Schr\"{o}dinger equation) like higher commuting
flows, conservation laws and local Hamiltonian structures. In Section~\ref%
{sec:reductions} we constructed new dispersive integrable systems. In
Section~\ref{sec:exception} separately we presented an exceptional case with
a dispersive chain but written in non-evolution form. In Section \ref%
{sec:quasi}) we investigated three dimensional linearly degenerate
quasilinear equations of a second order, we found explicit differential
constraints which allows to reduce these quasilinear equations to dispersive
integrable systems discussed in previous Sections. In Section~\ref%
{sec:dispersionless} we briefly derived a dispersionless limit of integrable
dispersive chains including their multi component dispersive reductions. In
Section \ref{sec:conclusion} we formulated a programme for a further
research.

\section{Dispersive Integrable Chains}

\label{sec:dispersive}

According to \cite{ShabatJNMP}, instead of the linear dependence $u(x,%
\mathbf{t},\lambda )=\lambda +u^{1}(x,\mathbf{t})$ (see (\ref{kdv})) and
quadratic dependence $u(x,\mathbf{t},\lambda )=\lambda ^{2}+\lambda u^{1}(x,%
\mathbf{t})+u^{2}(x,\mathbf{t})$ (see (\ref{kb})) below we consider ($%
M=1,2,...$)%
\begin{equation}
u(x,\mathbf{t},\lambda )=\lambda ^{M}\left( 1+\frac{u^{1}(x,\mathbf{t})}{%
\lambda }+\frac{u^{2}(x,\mathbf{t})}{\lambda ^{2}}+\frac{u^{3}(x,\mathbf{t})%
}{\lambda ^{3}}+...\right) ,  \label{laurent}
\end{equation}%
where $u^{k}$ are infinitely many unknown functions\footnote{%
this Laurent expansion was suggested in \cite{ShabatJNMP}; see other detail
in \cite{AdlerShabat}.}.

The substitution (\ref{laurent}) and the linear dependence $a^{(1)}=\lambda
+a_{1}(x,\mathbf{t})$ into (\ref{main}) yields $M$th integrable dispersive
chain\footnote{%
earlier in \cite{ShabatJNMP} these integrable dispersive chains were written
just in a compact symbolic form.}%
\begin{equation}
u_{t}^{k}=u_{x}^{k+1}-\frac{1}{2}u^{1}u_{x}^{k}-u^{k}u_{x}^{1}+\frac{1}{4}%
\delta _{M}^{k}u_{xxx}^{1},\text{ \ }k=1,2,...,  \label{mth}
\end{equation}%
where $\delta _{M}^{k}$ is the Kronecker delta and%
\begin{equation}
a_{1}=-\frac{1}{2}u^{1}.  \label{kon}
\end{equation}

\subsection{Higher Commuting flows}

Higher commuting flows of the Korteweg de Vries hierarchy are determined by
the linear spectral system%
\begin{equation*}
\psi _{xx}=(\lambda +u^{1})\psi ,\text{ \ }\psi _{t^{k}}=a^{(k)}\psi _{x}-%
\frac{1}{2}a_{x}^{(k)}\psi ,
\end{equation*}%
where
\begin{equation}
a^{(k)}=\lambda ^{k}+\overset{k}{\underset{m=1}{\sum }}a_{m}\lambda ^{k-m},
\label{ak}
\end{equation}%
and functions $a_{m}$ and $u^{1}$ depend on the \textquotedblleft
space\textquotedblright\ variable $x$ and infinitely many extra
\textquotedblleft time\textquotedblright\ variables $t^{k}$ (obviously, $%
t\equiv t^{1}$). Substitution (\ref{laurent}) and (\ref{ak}) into (\ref{main}%
) leads to higher commuting flows (here we also define $a_{0}=1$)%
\begin{equation}
u_{t^{s}}^{k}=\overset{s}{\underset{m=0}{\sum }}\left( u^{k+m}\partial
_{x}+\partial _{x}u^{k+m}-\frac{1}{2}\delta _{M}^{k+m}\partial
_{x}^{3}\right) a_{s-m},\text{ \ }s=1,2,...,  \label{com}
\end{equation}%
where all coefficients $a_{m}$ can be found iteratively from the linear
system\footnote{%
Linear system (\ref{con}) can be obtained formally from (\ref{com}), if to
fix $u^{0}=1$ and all other unknown functions $u^{-m}=0$ for $m=1,2,...$}
(here we also define $u^{0}=1$ and $u^{-m}=0$ for all $m=1,2,...$)%
\begin{equation}
\overset{s}{\underset{m=0}{\sum }}\left( u^{m-k}\partial _{x}+\partial
_{x}u^{m-k}-\frac{1}{2}\delta _{M}^{m-k}\partial _{x}^{3}\right) a_{s-m}=0,%
\text{ \ }k=0,1,...,s-1.  \label{con}
\end{equation}%
For instance,%
\begin{equation*}
a_{1}=-\frac{1}{2}u^{1},\text{ \ }a_{2}=-\frac{1}{2}u^{2}+\frac{3}{8}%
(u^{1})^{2}-\frac{1}{8}\delta _{M}^{1}u_{xx}^{1},
\end{equation*}%
\begin{equation*}
a_{3}=-\frac{1}{2}u^{3}+\frac{3}{4}u^{1}u^{2}-\frac{5}{16}(u^{1})^{3}+\frac{1%
}{32}\delta
_{M}^{1}(10u^{1}u_{xx}^{1}+5(u_{x}^{1})^{2}-u_{xxxx}^{1}-4u_{xx}^{2})-\frac{1%
}{8}\delta _{M}^{2}u_{xx}^{1},...
\end{equation*}%
Thus all higher commuting flows are written in an evolution form. For
instance, a first commuting flow to (\ref{mth}) is (here we identify $%
y\equiv t^{2}$)%
\begin{equation}
u_{y}^{k}=u_{x}^{k+2}-\frac{1}{2}u^{1}u_{x}^{k+1}+\left( -\frac{1}{2}u^{2}+%
\frac{3}{8}(u^{1})^{2}-\frac{1}{8}\delta _{M}^{1}u_{xx}^{1}\right)
u_{x}^{k}-u^{k+1}u_{x}^{1}  \label{c}
\end{equation}%
\begin{equation*}
+u^{k}\left( -u_{x}^{2}+\frac{3}{2}u^{1}u_{x}^{1}-\frac{1}{4}\delta
_{M}^{1}u_{xxx}^{1}\right) +\frac{1}{4}\delta _{M}^{k+1}u_{xxx}^{1}+\frac{1}{%
4}\delta _{M}^{k}\left( u_{xxx}^{2}-\frac{3}{4}[(u^{1})^{2}]_{xxx}+\frac{1}{4%
}\delta _{M}^{1}u_{xxxxx}^{1}\right) .
\end{equation*}

A generating function of higher commuting flows is determined by the choice
(instead of (\ref{ak}))%
\begin{equation}
a=\frac{\mathbf{a}(x,\mathbf{t},\zeta )}{\lambda -\zeta },  \label{ful}
\end{equation}%
where $\zeta $ is an arbitrary parameter. Indeed, substitution the
asymptotic expansion ($\zeta \rightarrow \infty $)%
\begin{equation}
\mathbf{a}(x,\mathbf{t},\zeta )=-\zeta \left( 1+\frac{a_{1}(x,\mathbf{t})}{%
\zeta }+\frac{a_{2}(x,\mathbf{t})}{\zeta ^{2}}+\frac{a_{3}(x,\mathbf{t})}{%
\zeta ^{3}}+...\right)  \label{ryad}
\end{equation}%
into (\ref{ful}) leads to%
\begin{equation*}
a=1+\frac{\lambda +a_{1}}{\zeta }+\frac{\lambda ^{2}+a_{1}\lambda +a_{2}}{%
\zeta ^{2}}+\frac{\lambda ^{3}+a_{1}\lambda ^{2}+a_{2}\lambda +a_{3}}{\zeta
^{3}}+...,
\end{equation*}%
where all polynomial expressions with respect to the spectral parameter $%
\lambda $ are precisely (\ref{ak}). The substitution (\ref{laurent}) and (%
\ref{ful}) into (\ref{main}) implies (let us remind that $u^{0}=1$)%
\begin{equation*}
u_{\tau (\zeta )}^{k+1}=\zeta u_{\tau (\zeta )}^{k}+2u^{k}\mathbf{a}_{x}+%
\mathbf{a}u_{x}^{k}-\frac{1}{2}\delta _{M}^{k}\mathbf{a}_{xxx},\text{ }%
k=0,1,...,
\end{equation*}%
where we introduced the special \textquotedblleft time\textquotedblright\
variable $\tau (\zeta )$ instead of $t$ in (\ref{main}) to emphasize that we
deal with a generating function of commuting flows. Taking into account that
$u_{\tau (\zeta )}^{1}=2\mathbf{a}_{x}$ and iteratively expressing all
others $u_{\tau (\zeta )}^{k}$ via higher derivatives of $u^{m}$ and $%
\mathbf{a}$, we come to a more explicit form%
\begin{equation*}
u_{\tau (\zeta )}^{k+1}=\overset{k}{\underset{m=0}{\sum }}\zeta
^{m}(u^{k-m}\partial _{x}+\partial _{x}u^{k-m}-\frac{1}{2}\delta
_{M}^{k-m}\partial _{x}^{3})\mathbf{a},\text{ }k=0,1,...,
\end{equation*}%
which again yields (\ref{com}), if to substitute (\ref{ryad}) together with
another formal expansion ($\zeta \rightarrow \infty $)%
\begin{equation*}
\partial _{\tau (\zeta )}=\partial _{t^{0}}+\frac{1}{\zeta }\partial
_{t^{1}}+\frac{1}{\zeta ^{2}}\partial _{t^{2}}+\frac{1}{\zeta ^{3}}\partial
_{t^{3}}+...
\end{equation*}%
Moreover, now we can identify $x\equiv t^{0}$.

\subsection{Conservation Laws}

The substitution%
\begin{equation*}
\psi =\exp \left( \int rdx\right)
\end{equation*}%
into (\ref{g}) yields%
\begin{equation}
r_{x}+r^{2}=u,\text{ \ }r_{t}=\left( ar-\frac{1}{2}a_{x}\right) _{x}.
\label{miura}
\end{equation}%
Thus $r(x,t,\lambda )$ is a generating function of conservation law
densities (with respect to the spectral parameter $\lambda $), while the
second equation of (\ref{miura}) is a generating function of conservation
laws.

The differential substitution ($\epsilon $ is an arbitrary constant)%
\begin{equation}
r=\frac{\varphi _{x}}{2\varphi }+\frac{\epsilon }{2\varphi }  \label{diff}
\end{equation}%
into the first equation of (\ref{miura}) yields%
\begin{equation}
2\varphi \varphi _{xx}-\varphi _{x}^{2}+\epsilon ^{2}=4u\varphi ^{2},
\label{kvadro}
\end{equation}%
which is nothing but a first integral of well-known equation (in the case of
the Korteweg de Vries equation)%
\begin{equation}
\varphi _{xxx}=4u\varphi _{x}+2u_{x}\varphi .  \label{third}
\end{equation}%
Indeed, the function $\varphi =\psi \psi ^{+}$ satisfies two linear equations%
\begin{equation}
\varphi _{xxx}=4u\varphi _{x}+2u_{x}\varphi ,\text{ \ }\varphi _{t}=a\varphi
_{x}-a_{x}\varphi ,  \label{k}
\end{equation}%
where (see (\ref{g})) $\psi $ and $\psi ^{+}$ are two linearly independent
solutions, which (obviously functionally dependent according to the
Wronskian relationship $\psi \psi _{x}^{+}-\psi ^{+}\psi _{x}=$const$\neq 0$%
) usually are determined by their asymptotic behavior ($\lambda \rightarrow
\infty $): $\psi \rightarrow \exp (\lambda ^{M/2}x),\psi ^{+}\rightarrow
\exp (-\lambda ^{M/2}x)$.

The second equation can be written in the conservative form%
\begin{equation}
\left( \frac{1}{\varphi }\right) _{t}=\left( \frac{a}{\varphi }\right) _{x}.
\label{zakon}
\end{equation}%
Thus the function $p=1/\varphi $ is a generating function of conservation
law densities, and the above equation is a generating function of
conservation laws. Taking into account (\ref{diff}), one can see that the
second equation in (\ref{miura}) is equivalent to the above generating
function of conservation laws (up to a total derivative).

\textbf{Theorem}: \textit{Dispersive chains} (\ref{mth}), (\ref{l}) \textit{%
also have an alternative generation function of conservation laws}%
\begin{equation}
(u^{\prime }\varphi )_{t}=\left( [2u+(\lambda +a_{1})u^{\prime }]\varphi -%
\frac{1}{2}\varphi _{xx}\right) _{x},  \label{gen}
\end{equation}%
\textit{where} $u^{\prime }\equiv u^{\prime }(\lambda )$, \textit{i.e.}%
\begin{equation*}
u=\lambda ^{M}+\overset{\infty }{\underset{k=1}{\sum }}\lambda ^{M-k}u^{k},%
\text{ \ }u^{\prime }=M\lambda ^{M-1}+\overset{\infty }{\underset{k=1}{\sum }%
}(M-k)\lambda ^{M-k-1}u^{k}.
\end{equation*}

\textbf{Proof}: We seek a generating function of conservation law densities
in the form $b\varphi $, where $b$ is a \textit{linear} expression with
respect field variables $u^{k}$, whose coefficients depend on the spectral
parameter $\lambda $ only. Differentiating this product $b\varphi $ with
respect to the independent variable \textquotedblleft $t$\textquotedblright\
and taking into account that (see the first equation in (\ref{k})) $%
u_{x}\varphi =(2u\varphi -\frac{1}{2}\varphi _{xx})_{x}$, we expect that the
flux of this generating function will be proportional to the function $%
\varphi $ and its second derivative $\varphi _{xx}$. A straightforward
computation yields that $b=u^{\prime }$, the flux is $[2u+(\lambda
+a_{1})u^{\prime }]\varphi -\frac{1}{2}\varphi _{xx}$ and\footnote{%
This equation can be obtained directly from (\ref{main}) by differentiation
with respect to the spectral parameter $\lambda $.}%
\begin{equation*}
u_{t}^{\prime }=(\lambda +a_{1})u_{x}^{\prime }+2u^{\prime }a_{1,x}+u_{x},
\end{equation*}%
where the function $a_{1}$ is defined by (\ref{kon}). The Theorem is proved.

Taking into account that (\ref{laurent}) we seek an asymptotic expansion of (%
\ref{third}) in the form (obviously, any solution of linear equation is
determined up to an arbitrary factor; thus without loss of generality and
for simplicity we choose the normalization to unity at infinity with respect
to the spectral parameter $\lambda $)%
\begin{equation}
\varphi (x,\mathbf{t},\lambda )=1+\frac{\varphi _{1}(x,\mathbf{t})}{\lambda }%
+\frac{\varphi _{2}(x,\mathbf{t})}{\lambda ^{2}}+\frac{\varphi _{3}(x,%
\mathbf{t})}{\lambda ^{3}}+...,\text{ \ }\lambda \rightarrow \infty .
\label{fi}
\end{equation}%
Then we obtain (here we define $\varphi _{0}=1$ and remind that $u^{0}=1$)%
\begin{equation*}
\frac{1}{2}\overset{\infty }{\underset{k=0}{\sum }}\frac{\varphi _{k+1,xxx}}{%
\lambda ^{M+k}}=\overset{\infty }{\underset{k=0}{\sum }}\frac{1}{\lambda ^{k}%
}\overset{k+1}{\underset{m=1}{\sum }}(2u^{k+1-m}\varphi _{m,x}+\varphi
_{k+1-m}u_{x}^{m}).
\end{equation*}%
Selecting factors of each degree of the spectral parameter $\lambda $, one
can see that expressions for coefficients $\varphi _{k}$ coincide with
expressions for coefficients $a_{k}$ in (\ref{con}), i.e. (\ref{fi}) becomes%
\begin{equation}
\varphi (x,\mathbf{t},\lambda )=1+\frac{a_{1}(x,\mathbf{t})}{\lambda }+\frac{%
a_{2}(x,\mathbf{t})}{\lambda ^{2}}+\frac{a_{3}(x,\mathbf{t})}{\lambda ^{3}}%
+...  \label{raz}
\end{equation}%
Thus substituting this asymptotic expansion into generating function (\ref%
{gen}), infinitely many conservation laws can be presented explicitly.

\subsection{Local Hamiltonian Structures}

\label{subsec:hamilton}

Construction of local multi-Hamiltonian structures for polynomial and
rational cases (\ref{racio}) was presented in \cite{lma}, \cite{antford},
\cite{antford2}.

A hierarchy of integrable dispersive chains (\ref{mth}) possesses \textit{%
infinitely} many local Hamiltonian structures, first two of them are ($%
s=1,2,...$)%
\begin{equation*}
u_{t^{s}}^{k}=\overset{s+1}{\underset{m=1}{\sum }}\left( u^{k+m-1}\partial
_{x}+\partial _{x}u^{k+m-1}-\frac{1}{2}\delta _{M}^{k+m-1}\partial
_{x}^{3}\right) \frac{\delta \mathbf{H}_{s+1}}{\delta u^{m}},\text{ }%
k=1,2,...;
\end{equation*}%
\begin{equation*}
u_{t^{s}}^{1}=-2\partial _{x}\frac{\delta \mathbf{H}_{s+2}}{\delta u^{1}},%
\text{ \ }u_{t^{s}}^{k}=\overset{s+2}{\underset{m=2}{\sum }}\left(
u^{k+m-2}\partial _{x}+\partial _{x}u^{k+m-2}-\frac{1}{2}\delta
_{M}^{k+m-2}\partial _{x}^{3}\right) \frac{\delta \mathbf{H}_{s+2}}{\delta
u^{m}},\text{ }k>1.
\end{equation*}

\textbf{Remark}: The first local Hamiltonian structure follows from (\ref%
{com}), where we utilized the observation%
\begin{equation}
a_{m}=\frac{\delta \mathbf{H}_{m+s}}{\delta u^{s}},\text{ }m=0,1,...;\text{ }%
s=1,2,...  \label{observ}
\end{equation}%
This is alternative approach for construction of local polynomial
conservation laws (cf. (\ref{miura}), (\ref{zakon}), (\ref{gen})). In such a
case all Hamiltonians can be found from above variation derivatives, for
instance%
\begin{equation*}
\mathbf{H}_{1}=\int u^{1}dx,\text{ \ }\mathbf{H}_{2}=\int \left( u^{2}-\frac{%
1}{4}(u^{1})^{2}\right) dx,
\end{equation*}%
\begin{equation*}
\mathbf{H}_{3}=\int \left( u^{3}-\frac{1}{2}u^{1}u^{2}+\frac{1}{8}%
(u^{1})^{3}+\frac{1}{16}\delta _{M}^{1}(u_{x}^{1})^{2}\right) dx,
\end{equation*}

\begin{equation*}
\begin{aligned} \mathbf{H}_{4} &=\int \biggl(
u^{4}-\frac{1}{2}u^{1}u^{3}-\frac{1}{4}(u^{2})^{2}+%
\frac{3}{8}(u^{1})^{2}u^{2}-\frac{5}{64}(u^{1})^{4} \\ &+\frac{1}{32}\delta
_{M}^{1}\left(
-5u^{1}(u_{x}^{1})^{2}-\frac{1}{2}(u_{xx}^{1})^{2}+4u_{x}^{1}u_{x}^{2}%
\right) +\frac{1}{16}\delta _{M}^{2}(u_{x}^{1})^{2}\biggr) dx,...
\end{aligned}
\end{equation*}%
All other higher local Hamiltonian structures can be constructed utilizing
the relationship (\ref{observ}), i.e.%
\begin{equation*}
\frac{\delta \mathbf{H}_{s}}{\delta u^{m}}=\frac{\delta \mathbf{H}_{s+k}}{%
\delta u^{m+k}},\text{ \ \ }m,s,k=1,2,...
\end{equation*}%
For instance, the third local Hamiltonian structure is ($s=1,2,...$)%
\begin{equation*}
u_{t^{s}}^{1}=-2\partial _{x}\frac{\delta \mathbf{H}_{s+3}}{\delta u^{2}},%
\text{ \ }u_{t^{s}}^{2}=-2\partial _{x}\frac{\delta \mathbf{H}_{s+3}}{\delta
u^{1}}-\left( u^{1}\partial _{x}+\partial _{x}u^{1}-\frac{1}{2}\delta
_{M}^{1}\partial _{x}^{3}\right) \frac{\delta \mathbf{H}_{s+3}}{\delta u^{2}}%
,
\end{equation*}%
\begin{equation*}
u_{t^{s}}^{k}=\overset{s+3}{\underset{m=3}{\sum }}\left( u^{k+m-3}\partial
_{x}+\partial _{x}u^{k+m-3}-\frac{1}{2}\delta _{M}^{k+m-3}\partial
_{x}^{3}\right) \frac{\delta \mathbf{H}_{s+3}}{\delta u^{m}},\text{ }k>2.
\end{equation*}

\section{Multi Component Reductions}

\label{sec:reductions}

Theory of multi-component semi-Hamiltonian hydrodynamic reductions of
integrable hydrodynamic chains was build in \cite{gibtsar} and applied in
several papers \cite{ferap}, \cite{maksjmp} (see also \cite{mas2}, \cite%
{makschain}). Corresponding theory of multi-component integrable \textit{%
dispersive} reductions of integrable \textit{dispersive} chains does not
exist at this moment. Nevertheless infinitely many multi-component
integrable dispersive systems extracted from (\ref{mth}) are presented in
this Section.

\subsection{Elementary Reductions}

Obviously for any natural number $N\geqslant M$ the reduction $u^{N+1}=0$ of
$M$th dispersive chain (\ref{mth}) leads to $N$ component integrable
dispersive systems:

\textbf{1}. $N=M=1$, this is the Korteweg de Vries equation (\ref{kdv});

\textbf{2}. $N=2,M=1$, this is the Ito system%
\begin{equation}
u_{t}^{1}=u_{x}^{2}-\frac{3}{2}u^{1}u_{x}^{1}+\frac{1}{4}u_{xxx}^{1},\text{
\ }u_{t}^{2}=-\frac{1}{2}u^{1}u_{x}^{2}-u^{2}u_{x}^{1};  \label{ito}
\end{equation}

\textbf{3}. $N>2,M=1$,%
\begin{equation*}
u_{t}^{1}=u_{x}^{2}-\frac{3}{2}u^{1}u_{x}^{1}+\frac{1}{4}u_{xxx}^{1},
\end{equation*}%
\begin{equation*}
u_{t}^{k}=u_{x}^{k+1}-\frac{1}{2}u^{1}u_{x}^{k}-u^{k}u_{x}^{1},\text{ \ }%
k=2,...,N-1,
\end{equation*}%
\begin{equation*}
u_{t}^{N}=-\frac{1}{2}u^{1}u_{x}^{N}-u^{N}u_{x}^{1};
\end{equation*}

\textbf{4}. $N=M>1$, (if $N=M=2$, this is the Kaup--Boussinesq equation, see
(\ref{kb}))%
\begin{equation*}
u_{t}^{k}=u_{x}^{k+1}-\frac{1}{2}u^{1}u_{x}^{k}-u^{k}u_{x}^{1},\text{ \ }%
k=1,2,...,N-1,
\end{equation*}%
\begin{equation*}
u_{t}^{N}=-\frac{1}{2}u^{1}u_{x}^{N}-u^{N}u_{x}^{1}+\frac{1}{4}u_{xxx}^{1};
\end{equation*}

\textbf{5}. $N=M+1,M>1$,%
\begin{equation*}
u_{t}^{k}=u_{x}^{k+1}-\frac{1}{2}u^{1}u_{x}^{k}-u^{k}u_{x}^{1},\text{ }%
k=1,2,...,N-2,
\end{equation*}%
\begin{equation*}
u_{t}^{N-1}=u_{x}^{N}-\frac{1}{2}u^{1}u_{x}^{N-1}-u^{N-1}u_{x}^{1}+\frac{1}{4%
}u_{xxx}^{1},
\end{equation*}%
\begin{equation*}
u_{t}^{N}=-\frac{1}{2}u^{1}u_{x}^{N}-u^{N}u_{x}^{1}.
\end{equation*}

\textbf{6}. $N>M+1,M>1$,%
\begin{equation*}
u_{t}^{k}=u_{x}^{k+1}-\frac{1}{2}u^{1}u_{x}^{k}-u^{k}u_{x}^{1},\text{ }%
k=1,...,M,
\end{equation*}%
\begin{equation*}
u_{t}^{M}=u_{x}^{M+1}-\frac{1}{2}u^{1}u_{x}^{M}-u^{M}u_{x}^{1}+\frac{1}{4}%
u_{xxx}^{1},
\end{equation*}%
\begin{equation*}
u_{t}^{k}=u_{x}^{k+1}-\frac{1}{2}u^{1}u_{x}^{k}-u^{k}u_{x}^{1},\text{ \ }%
k=M+1,...,N-1,
\end{equation*}%
\begin{equation*}
u_{t}^{N}=-\frac{1}{2}u^{1}u_{x}^{N}-u^{N}u_{x}^{1}.
\end{equation*}

\subsection{Rational Constraints with Movable Singularities}

\label{subsec:racio}

Now we consider more complicated $N$ component reductions ($M=1,2,...$)%
\begin{equation}
u(x,\mathbf{t},\lambda )=\frac{\lambda ^{M+K}+\lambda ^{M+K-1}v_{M+K-1}(x,%
\mathbf{t})+...+\lambda v_{1}(x,\mathbf{t})+v_{0}(x,\mathbf{t})}{\lambda
^{K}+\lambda ^{K-1}w_{K-1}(x,\mathbf{t})+...+\lambda w_{1}(x,\mathbf{t}%
)+w_{0}(x,\mathbf{t})},\text{ \ }K=0,1,...  \label{i}
\end{equation}%
Suppose for simplicity that all roots of these two polynomials are pairwise
distinct. Then the substitution%
\begin{equation}
u(x,\mathbf{t},\lambda )=\frac{\overset{M+K}{\underset{m=1}{\dprod }}%
(\lambda -s^{m}(x,\mathbf{t}))}{\overset{K}{\underset{k=1}{\dprod }}(\lambda
-r^{k}(x,\mathbf{t}))}  \label{rac}
\end{equation}%
into (\ref{main}) together with the linear dependence $a^{(1)}=\lambda
+a_{1}(x,\mathbf{t})$ yields new multi-component integrable dispersive
systems%
\begin{equation}
r_{t}^{k}=(r^{k}+a_{1})r_{x}^{k},\text{ \ }s_{t}^{i}=(s^{i}+a_{1})s_{x}^{i}+%
\frac{1}{2}\frac{\overset{K}{\underset{k=1}{\dprod }}(s^{i}-r^{k})}{\underset%
{m\neq i}{\dprod }(s^{i}-s^{m})}a_{1,xxx},  \label{disp}
\end{equation}%
where%
\begin{equation}
a_{1}=\frac{1}{2}\left( \overset{M+K}{\underset{m=1}{\sum }}s^{m}-\overset{K}%
{\underset{k=1}{\sum }}r^{k}\right) .  \label{a1}
\end{equation}

\textbf{Remark}: In the particular case $r^{k}=$const, the ansatz (\ref{i}) (%
$\epsilon _{k}$ are symmetric functions with respect to $r^{k}$ according to
the Vi\`{e}te theorem)%
\begin{equation*}
u(x,\mathbf{t},\lambda )=\frac{\lambda ^{M+K}+\lambda ^{M+K-1}v_{M+K-1}(x,%
\mathbf{t})+...+\lambda v_{1}(x,\mathbf{t})+v_{0}(x,\mathbf{t})}{\lambda
^{K}+\epsilon _{K-1}\lambda ^{K-1}+...+\epsilon _{1}\lambda +\epsilon _{0}}
\end{equation*}%
was investigated in \cite{antford2}. Then $2K+M$ component system (\ref{disp}%
) reduces to the form%
\begin{equation*}
s_{t}^{i}=(s^{i}+a_{1})s_{x}^{i}+\frac{1}{2}\frac{\overset{K}{\underset{k=1}{%
\dprod }}(s^{i}-r^{k})}{\underset{m\neq i}{\dprod }(s^{i}-s^{m})}a_{1,xxx},%
\text{ }i=1,2,...,M+K,
\end{equation*}%
where $r^{k}=$const. Thus all integrable systems considered in \cite%
{antford2} also can be written in the above symmetric form. For instance,
the Kaup--Boussinesq system (\ref{kb}) becomes%
\begin{equation}
s_{t}^{1}=\frac{1}{2}(3s^{1}+s^{2})s_{x}^{1}+\frac{(s^{1}+s^{2})_{xxx}}{%
4(s^{1}-s^{2})},\text{ }s_{t}^{2}=\frac{1}{2}(s^{1}+3s^{2})s_{x}^{2}-\frac{%
(s^{1}+s^{2})_{xxx}}{4(s^{1}-s^{2})};  \label{j}
\end{equation}%
the Ito system (\ref{ito}) takes the form%
\begin{equation*}
s_{t}^{1}=\frac{1}{2}(3s^{1}+s^{2})s_{x}^{1}+\frac{s^{1}(s^{1}+s^{2})_{xxx}}{%
4(s^{1}-s^{2})},\text{ }s_{t}^{2}=\frac{1}{2}(s^{1}+3s^{2})s_{x}^{2}-\frac{%
s^{2}(s^{1}+s^{2})_{xxx}}{4(s^{1}-s^{2})}.
\end{equation*}

\subsection{Negative Flows}

Substitution the expansion ($\zeta \rightarrow 0$)%
\begin{equation*}
\mathbf{a}(x,\mathbf{t},\zeta )=-a_{-1}(x,\mathbf{t})-\zeta a_{-2}(x,\mathbf{%
t})-\zeta ^{2}a_{-3}(x,\mathbf{t})+...
\end{equation*}%
into (\ref{ful}) yields (cf. (\ref{ryad}))%
\begin{equation*}
a=-\frac{a_{-1}}{\lambda }-\left( \frac{a_{-2}}{\lambda }+\frac{a_{-1}}{%
\lambda ^{2}}\right) \zeta -\left( \frac{a_{-3}}{\lambda }+\frac{a_{-2}}{%
\lambda ^{2}}+\frac{a_{-1}}{\lambda ^{3}}\right) \zeta ^{2}-...
\end{equation*}%
Substitution this expansion and the rational ansatz (\ref{rac}) into (\ref%
{main}) implies infinitely many lower (negative) flows of the integrable
hierarchies, whose first nontrivial (positive) members are (\ref{disp}). For
instance, the choice%
\begin{equation}
a=-\frac{a_{-1}}{\lambda }  \label{nega}
\end{equation}%
determines non-evolution system%
\begin{equation}
r_{t^{-1}}^{k}=-\frac{a_{-1}}{r^{k}}r_{x}^{k},\text{ \ }s_{t^{-1}}^{i}=-%
\frac{a_{-1}}{s^{i}}s_{x}^{i}-\frac{1}{2s^{i}}\frac{\overset{K}{\underset{k=1%
}{\dprod }}(s^{i}-r^{k})}{\underset{m\neq i}{\dprod }(s^{i}-s^{m})}%
a_{-1,xxx},  \label{nonloc}
\end{equation}%
where%
\begin{equation}
a_{1,t^{-1}}=a_{-1,x}.  \label{aa}
\end{equation}%
Taking into account (\ref{a1}), one can derive an ordinary differential
equation ($\tilde{\xi}$ is an integration constant) on the function $a_{-1}$%
, i.e. if any $r^{k}\neq 0$, then%
\begin{equation}
\frac{(-1)^{M}}{2}\left( \frac{a_{-1,xx}}{a_{-1}}-\frac{1}{2}\frac{%
a_{-1,x}^{2}}{a_{-1}^{2}}\right) +\frac{\tilde{\xi}}{a_{-1}^{2}}=\frac{%
\overset{M+K}{\underset{i=1}{\dprod }}s^{i}}{\overset{K}{\underset{k=1}{%
\dprod }}r^{k}}.  \label{ogra}
\end{equation}%
If, for instance, $r^{1}=0$, then non-evolution system (\ref{nonloc})
becomes the \textit{evolution} system ($k=2,3,...,K;i=1,2,...,M+K$), where ($%
\bar{\xi}$ is an integration constant)%
\begin{equation}
a_{-1}=\bar{\xi}\left( \frac{\overset{K}{\underset{k=2}{\dprod }}r^{k}}{%
\overset{M+K}{\underset{m=1}{\dprod }}s^{m}}\right) ^{1/2}.  \label{okra}
\end{equation}

\section{The Exceptional Case}

\label{sec:exception}

The concept of $M$th dispersive integrable chain can be extended to the case
$M=0$ (see (\ref{laurent})). Indeed, the compatibility condition $(\psi
_{xx})_{t}=(\psi _{t})_{xx}$\ of the linear system%
\begin{equation*}
\psi _{xx}=\left( 1+\frac{u^{1}}{\lambda }+\frac{u^{2}}{\lambda ^{2}}%
+...\right) \psi ,\text{ \ }\psi _{t}=(\lambda +a_{1})\psi _{x}-\frac{1}{2}%
a_{1,x}\psi
\end{equation*}%
yields the \textit{zeroth} dispersive integrable chain%
\begin{equation}
u_{t}^{k}=u_{x}^{k+1}+a_{1}u_{x}^{k}+2u^{k}a_{1,x},\text{ \ }k=1,2,...,
\label{l}
\end{equation}%
where ($\xi $ is an arbitrary constant)%
\begin{equation}
\frac{1}{2}a_{1,xx}-2a_{1}=u^{1}+\xi .  \label{cel}
\end{equation}%
Since the function $a_{1}$ cannot be expressed via the function $u_{1}$ and
a finite number of its derivatives, this zeroth dispersive chain is not an
evolution system (in comparison with dispersive chains (\ref{mth}) with $M>0$%
).

For this particular case we omit consideration of conservation laws,
commuting flows and Hamiltonian structures, because such an investigation
can be made exactly as in previous Sections. Here we just mention most
important non-evolution reductions:

\textbf{1}. $N=1$, the Camassa--Holm equation:%
\begin{equation}
u_{t}^{1}=a_{1}u_{x}^{1}+2u^{1}a_{1,x},\text{ \ }\frac{1}{2}%
a_{1,xx}-2a_{1}=u^{1}+\xi ;  \label{ch}
\end{equation}

\textbf{2}. $N>1$, the multi-component generalization of Camassa--Holm
equation:%
\begin{equation}
u_{t}^{k}=u_{x}^{k+1}+a_{1}u_{x}^{k}+2u^{k}a_{1,x},\text{ \ }k=1,2,...,N-1,
\label{ch2}
\end{equation}%
\begin{equation*}
u_{t}^{N}=a_{1}u_{x}^{N}+2u^{N}a_{1,x},\text{ \ }\frac{1}{2}%
a_{1,xx}-2a_{1}=u^{1}+\xi .
\end{equation*}

\textbf{3}. Suppose for simplicity that all roots of two polynomials in (\ref%
{i}) are pairwise distinct. Then the substitution%
\begin{equation}
u(x,\mathbf{t},\lambda )=\frac{\overset{K}{\underset{m=1}{\dprod }}(\lambda
-s^{m}(x,\mathbf{t}))}{\overset{K}{\underset{k=1}{\dprod }}(\lambda -r^{k}(x,%
\mathbf{t}))}  \label{racek}
\end{equation}%
into (\ref{main}) yields multi-component non-evolution systems (cf. (\ref%
{disp}))%
\begin{equation*}
r_{t}^{k}=(r^{k}+a_{1})r_{x}^{k},\text{ \ }s_{t}^{i}=(s^{i}+a_{1})s_{x}^{i}+%
\frac{1}{2}\frac{\overset{K}{\underset{k=1}{\dprod }}(s^{i}-r^{k})}{\underset%
{m\neq i}{\dprod }(s^{i}-s^{m})}a_{1,xxx},
\end{equation*}%
where $a_{1}$ is determined by ($\xi $ is an arbitrary constant, see (\ref%
{cel}))%
\begin{equation}
2a_{1}-\frac{1}{2}a_{1,xx}=\overset{K}{\underset{m=1}{\sum }}s^{m}-\overset{K%
}{\underset{k=1}{\sum }}r^{k}-\xi .  \label{aksi}
\end{equation}

In this case lower commuting flows are determined by (\ref{nega}) and (\ref%
{racek}). Then (\ref{main}) leads to (cf. (\ref{nonloc}))%
\begin{equation}
r_{t^{-1}}^{k}=-\frac{a_{-1}}{r^{k}}r_{x}^{k},\text{ \ }s_{t^{-1}}^{i}=-%
\frac{a_{-1}}{s^{i}}s_{x}^{i}-\frac{1}{2s^{i}}\frac{\overset{K}{\underset{k=1%
}{\dprod }}(s^{i}-r^{k})}{\underset{m\neq i}{\dprod }(s^{i}-s^{m})}%
a_{-1,xxx},  \label{lower}
\end{equation}%
where the function $a_{-1}$ satisfies (\ref{aa}). Taking into account (\ref%
{aksi}), one can derive an ordinary differential equation ($\tilde{\xi}$ is
an integration constant) on the function $a_{-1}$, i.e. if any $r^{k}\neq 0$%
, then (cf. (\ref{ogra}))%
\begin{equation*}
\frac{1}{2}(-1)^{M}\left( \frac{a_{-1,xx}}{a_{-1}}-\frac{1}{2}\frac{%
a_{-1,x}^{2}}{a_{-1}^{2}}\right) +\frac{\tilde{\xi}}{a_{-1}^{2}}=\frac{%
\overset{K}{\underset{m=1}{\dprod }}s^{m}}{\overset{K}{\underset{k=1}{\dprod
}}r^{k}}.
\end{equation*}%
If, for instance, $r^{1}=0$, then non-evolution system (\ref{lower}) becomes
the \textit{evolution} system%
\begin{equation*}
r_{t^{-1}}^{k}=-\frac{a_{-1}}{r^{k}}r_{x}^{k},\text{ \ }s_{t^{-1}}^{i}=-%
\frac{a_{-1}}{s^{i}}s_{x}^{i}-\frac{1}{2}\frac{\overset{K}{\underset{k=2}{%
\dprod }}(s^{i}-r^{k})}{\underset{m\neq i}{\dprod }(s^{i}-s^{m})}a_{-1,xxx}
\end{equation*}%
($k=2,3,...,K;i=1,2,...,K$), where ($\bar{\xi}$ is an integration constant,
cf. (\ref{okra}))%
\begin{equation*}
a_{-1}=\bar{\xi}\left( \frac{\overset{K}{\underset{k=2}{\dprod }}r^{k}}{%
\overset{K}{\underset{m=1}{\dprod }}s^{m}}\right) ^{1/2}.
\end{equation*}

\section{Three Dimensional Linearly Degenerate Quasilinear Equations}

\label{sec:quasi}

Hierarchies of integrable dispersive chains (\ref{mth}), (\ref{com}) have
generating functions of conservation laws%
\begin{equation*}
p_{t^{k}}=(a^{(k)}p)_{x},
\end{equation*}%
where $p=1/\varphi $ (see (\ref{zakon})). The compatibility conditions $%
(p_{t^{k}})_{t^{m}}=(p_{t^{m}})_{t^{k}}$ must be fulfilled, because
corresponding dispersive integrable chains (and their multi component
reductions) commute to each other. For instance, consistency of two first
such equations (we remind that $x=t^{0},t=t^{1},y=t^{2}$)%
\begin{equation}
p_{t}=[(\lambda +a_{1})p]_{x},\text{ \ }p_{y}=[(\lambda ^{2}+a_{1}\lambda
+a_{2})p]_{x}  \label{pi}
\end{equation}%
leads to the three dimensional quasilinear system%
\begin{equation}
a_{1,t}=a_{2,x},\text{ \ }a_{1}a_{2,x}+a_{1,y}=a_{2}a_{1,x}+a_{2,t},
\label{three}
\end{equation}%
whose integrability by the method of hydrodynamic reductions was
investigated in \cite{maksjmp}. This three dimensional quasilinear system
also belongs to the class of such integrable systems which called \textit{%
linearly degenerate}. This means that any of these systems admits at least
one $N$ component two dimensional hydrodynamic reduction ($N$ must be
arbitrary) which is \textit{linearly degenerate} (see detail in \cite%
{Makslin}, \cite{Fer91}). Moreover in such a case all hydrodynamic
reductions can be completely described (see, for instance, \cite{maksjmp}).
They are parameterised by $N$ arbitrary functions of a single variable.

In this Section we formulate the following

\textbf{Statement}: \textit{Three dimensional quasilinear system} (\ref%
{three}) \textit{possesses infinitely many differential constraints} $a_{1}(%
\mathbf{u},\mathbf{u}_{x},...),a_{2}(\mathbf{u},\mathbf{u}_{x},...)$,
\textit{where field variables} $u^{k}$ \textit{are solutions of dispersive
integrable systems determined by linear spectral problem} (\ref{g}) \textit{%
and described in Sections} \textbf{3} \textit{and} \textbf{4}.

Already in two first papers \cite{mas} a concept of differential constraints
was introduced for the hydrodynamic chain%
\begin{equation}
a_{k,t}=a_{k+1,x}+a_{1}a_{k,x}-a_{k}a_{1,x},\text{ }k=1,2,...,  \label{a}
\end{equation}%
which can be obtained from the second equation in (\ref{g}), where one
should substitute the linear ansatz $a^{(1)}=\lambda +a_{1}$ instead of $a$
and formal expansion (\ref{raz}) instead of $\varphi $. L. Martinez Alonso
and A.B. Shabat also considered such remarkable examples like the Korteweg
de Vries equation and the nonlinear Schr\"{o}dinger equation.

The first commuting flow to (\ref{a})%
\begin{equation}
a_{k,y}=a_{k+2,x}+a_{1}a_{k+1,x}-a_{k+1}a_{1,x}+a_{2}a_{k,x}-a_{k}a_{2,x},%
\text{ }k=1,2,...  \label{b}
\end{equation}%
can be obtained from the second equation in (\ref{g}), where one should
substitute the quadratic ansatz $a^{(2)}=\lambda ^{2}+a_{1}\lambda +a_{2}$
instead of $a$ and formal expansion (\ref{raz}) instead of $\varphi $.
Taking first two equations $%
a_{1,t}=a_{2,x},a_{2,t}=a_{3,x}+a_{1}a_{2,x}-a_{2}a_{1,x}$ from (\ref{a}),
the first equation $a_{1,y}=a_{3,x}$ from (\ref{b}), and excluding $a_{3,x}$%
, one can obtain three dimensional quasilinear system (\ref{three}).

\textbf{Remark}: \textit{Hydrodynamic} chain (\ref{a}) under differential
invertible polynomial substitutions $a_{k}(\mathbf{u},\mathbf{u}_{x},...)$
from (\ref{con}) transforms into \textit{dispersive} chains (\ref{mth}),
where $M$ is any natural number. This means that any $M$th dispersive chain
is \textit{equivalent} to another $K$th dispersive chain by appropriate
invertible differential substitutions (also including the exceptional case $%
M=0$). However, corresponding differential reductions are not equivalent to
each other.

Substitution the quadratic dependence $a^{(2)}=\lambda ^{2}+\lambda
a_{1}+a_{2}$ into (\ref{main}) together with ansatz (\ref{rac}) yields first
commuting flow to (\ref{disp})%
\begin{equation*}
r_{y}^{k}=(a_{2}+a_{1}r^{k}+(r^{k})^{2})r_{x}^{k},
\end{equation*}%
\begin{equation}
s_{y}^{i}=(a_{2}+a_{1}s^{i}+(s^{i})^{2})s_{x}^{i}+\frac{1}{2}\frac{\overset{K%
}{\underset{k=1}{\dprod }}(s^{i}-r^{k})}{\underset{m\neq i}{\dprod }%
(s^{i}-s^{m})}(s^{i}a_{1,xxx}+a_{2,xxx}),  \label{diss}
\end{equation}%
where $a_{1}$ is determined by (\ref{a1}) and
\begin{equation}
a_{2}=\frac{1}{4}\overset{M+K}{\underset{m=1}{\sum }}(s^{m})^{2}-\frac{1}{4}%
\overset{K}{\underset{k=1}{\sum }}(r^{k})^{2}+\frac{1}{2}a_{1}^{2}+\frac{1}{4%
}\delta _{M}^{1}a_{1,xx}.  \label{a2}
\end{equation}%
Thus in this Section we selected dispersive constraints (\ref{a1}), (\ref{a2}%
) of three dimensional quasilinear system (\ref{three}), where $M+K$
functions $s^{i}(x,t,y)$ and $K$ functions $r^{k}(x,t,y)$ are solutions of $%
2K+M$ component commuting dispersive integrable systems (\ref{disp}) and (%
\ref{diss}).

Now we present first four important examples of finite component
differential reductions of three dimensional quasilinear system (\ref{three}%
).

\textbf{1}. The Korteweg de Vries equation ($N=M=1$). In this case~commuting
dispersive chains (\ref{mth}) and (\ref{c}) reduce to (see (\ref{kdv}))%
\begin{equation*}
u_{t}^{1}=\left( \frac{1}{4}u_{xx}^{1}-\frac{3}{4}(u^{1})^{2}\right) _{x},
\end{equation*}%
and its first commuting flow%
\begin{equation*}
u_{y}^{1}=\left( \frac{5}{8}(u^{1})^{3}-\frac{5}{16}(u_{x}^{1})^{2}-\frac{5}{%
8}u^{1}u_{xx}^{1}+\frac{1}{16}u_{xxxx}^{1}\right) _{x}.
\end{equation*}%
Substitution (see (\ref{con}))%
\begin{equation*}
a_{1}=-\frac{1}{2}u^{1},\text{ \ }a_{2}=\frac{3}{8}(u^{1})^{2}-\frac{1}{8}%
u_{xx}^{1}
\end{equation*}%
into three dimensional quasilinear system (\ref{three}) leads to an identity.

\textbf{2}. The Ito system ($N=2,M=1$). In this case~commuting dispersive
chains (\ref{mth}) and (\ref{c}) reduce to (see (\ref{ito}))%
\begin{equation*}
u_{t}^{1}=u_{x}^{2}-\frac{3}{2}u^{1}u_{x}^{1}+\frac{1}{4}u_{xxx}^{1},\text{
\ }u_{t}^{2}=-\frac{1}{2}u^{1}u_{x}^{2}-u^{2}u_{x}^{1},
\end{equation*}%
and its first commuting flow%
\begin{equation*}
u_{y}^{1}=-\frac{3}{2}u^{1}u_{x}^{2}+\left( -\frac{3}{2}u^{2}+\frac{15}{8}%
(u^{1})^{2}-\frac{1}{8}u_{xx}^{1}\right) u_{x}^{1}
\end{equation*}%
\begin{equation*}
-\frac{1}{4}u^{1}u_{xxx}^{1}+\frac{1}{4}u_{xxx}^{2}-\frac{3}{16}%
[(u^{1})^{2}]_{xxx}+\frac{1}{16}u_{xxxxx}^{1},
\end{equation*}%
\begin{equation*}
u_{y}^{2}=\left( -\frac{3}{2}u^{2}+\frac{3}{8}(u^{1})^{2}-\frac{1}{8}%
u_{xx}^{1}\right) u_{x}^{2}+\frac{3}{2}u^{2}u^{1}u_{x}^{1}-\frac{1}{4}%
u^{2}u_{xxx}^{1}.
\end{equation*}%
Substitution (see (\ref{con}))%
\begin{equation*}
a_{1}=-\frac{1}{2}u^{1},\text{ \ }a_{2}=-\frac{1}{2}u^{2}+\frac{3}{8}%
(u^{1})^{2}-\frac{1}{8}u_{xx}^{1},
\end{equation*}%
into three dimensional quasilinear system (\ref{three}) leads to an identity.

\textbf{3}. The Kaup--Boussinesq equation ($N=M=2$). In this case~commuting
dispersive chains (\ref{mth}) and (\ref{c}) reduce to (see (\ref{kb}))%
\begin{equation*}
u_{t}^{1}=u_{x}^{2}-\frac{3}{2}u^{1}u_{x}^{1},\text{ \ }u_{t}^{2}=\frac{1}{4}%
u_{xxx}^{1}-u^{2}u_{x}^{1}-\frac{1}{2}u^{1}u_{x}^{2},
\end{equation*}%
and its first commuting flow%
\begin{equation*}
u_{y}^{1}=-\frac{3}{2}u^{1}u_{x}^{2}-\frac{3}{2}u^{2}u_{x}^{1}+\frac{15}{8}%
(u^{1})^{2}u_{x}^{1}+\frac{1}{4}u_{xxx}^{1},
\end{equation*}%
\begin{equation*}
u_{y}^{2}=-\frac{3}{2}u^{2}u_{x}^{2}+\frac{3}{8}(u^{1})^{2}u_{x}^{2}+\frac{3%
}{2}u^{2}u^{1}u_{x}^{1}+\frac{1}{4}u_{xxx}^{2}-\frac{3}{16}%
[(u^{1})^{2}]_{xxx}.
\end{equation*}%
Substitution (see (\ref{con}))%
\begin{equation*}
a_{1}=-\frac{1}{2}u^{1},\text{ \ }a_{2}=-\frac{1}{2}u^{2}+\frac{3}{8}%
(u^{1})^{2}
\end{equation*}%
into three dimensional quasilinear system (\ref{three}) leads to an identity.

\textbf{4}. The Camassa--Holm equation ($N=1,M=0$). Substitution expressions
($\xi _{1},\xi _{2}$ are arbitrary parameters)%
\begin{equation*}
a_{1}=-\frac{1}{2}\left( 1-\frac{1}{4}\partial _{x}^{2}\right)
^{-1}(u^{1}+\xi _{1}),
\end{equation*}%
\begin{equation*}
a_{2}=\left( 1-\frac{1}{4}\partial _{x}^{2}\right) ^{-1}\left( -\frac{1}{4}%
a_{1}a_{1,xx}-\frac{1}{8}a_{1,x}^{2}+\frac{3}{2}a_{1}^{2}+\xi _{1}a_{1}+\xi
_{2}\right)
\end{equation*}%
into three dimensional quasilinear system (\ref{three}) leads to an
identity, if the function $u^{1}(x,t,y)$ is a solution of the pair of
nonlocal equations (see (\ref{ch}))%
\begin{equation*}
u_{t}^{1}=a_{1}u_{x}^{1}+2u^{1}a_{1,x},\text{ \ }%
u_{y}^{1}=2u^{1}a_{2,x}+a_{2}u_{x}^{1}.
\end{equation*}

\subsection{Negative Flows}

Now we consider another asymptotic expansion of the function (cf. (\ref{ful}%
), (\ref{ryad}))%
\begin{equation}
\mathbf{a}(x,\mathbf{t},\zeta )=-a_{-1}-\zeta a_{-2}-\zeta ^{2}a_{-3}-...,%
\text{ \ }\zeta \rightarrow 0  \label{ya}
\end{equation}%
Then consistency of two generating functions of conservation laws (cf. (\ref%
{pi}))%
\begin{equation*}
p_{t^{-1}}=-\frac{1}{\lambda }(a_{-1}p)_{x},\text{ \ }p_{t}=[(\lambda
+a_{1})p]_{x}
\end{equation*}%
implies the second three dimensional quasilinear system (cf. (\ref{three}))%
\begin{equation}
a_{1,t^{-1}}=a_{-1,x},\text{ \ }a_{-1,t}=a_{1}a_{-1,x}-a_{-1}a_{1,x};
\label{dva}
\end{equation}%
consistency of two generating functions of conservation laws%
\begin{equation}
p_{t^{-1}}=-\left( \frac{a_{-1}}{\lambda }p\right) _{x},\text{ \ }%
p_{t^{-2}}=-\left[ \left( \frac{a_{-2}}{\lambda }+\frac{a_{-1}}{\lambda ^{2}}%
\right) p\right] _{x}  \label{negadva}
\end{equation}%
leads to the third three dimensional quasilinear system%
\begin{equation}
a_{-1,t^{-2}}=a_{-2,t^{-1}},\text{ \ }%
a_{-1,t^{-1}}=a_{-2}a_{-1,x}-a_{-1}a_{-2,x}.  \label{odin}
\end{equation}

\textbf{Remark}: Taking into account that $p=1/\varphi $, substitution
formal expansion (\ref{raz}) instead of $\varphi $ into (\ref{negadva})
yields two commuting hydrodynamic chains\footnote{%
let us remind that $a_{0}=1$.} to (\ref{a}), (\ref{b})%
\begin{equation}
a_{k+1,t^{-1}}=a_{k}a_{-1,x}-a_{-1}a_{k,x},\text{ }k=0,1,2,...  \label{f}
\end{equation}%
\begin{equation}
a_{k+2,t^{-2}}=a_{k+1}a_{-2,x}-a_{-2}a_{k+1,x}+a_{k}a_{-1,x}-a_{-1}a_{k,x},%
\text{ }k=-1,0,1,...  \label{h}
\end{equation}%
Moreover, all above hydrodynamic chains can be extended on negative values
of the index $k$ (see, for instance, \cite{maksjmp} and \cite{mas2}). Thus,
taking the first \textquotedblleft negative\textquotedblright\ equation of
hydrodynamic chain (\ref{a})%
\begin{equation*}
a_{-1,t}=a_{1}a_{-1,x}-a_{-1}a_{1,x}
\end{equation*}%
together with the first equation of hydrodynamic chain (\ref{f}) $%
a_{1,t^{-1}}=a_{-1,x}$, one can obtain three dimensional quasilinear system (%
\ref{dva}); taking the first two \textquotedblleft
negative\textquotedblright\ equations of hydrodynamic chain (\ref{f})%
\begin{equation*}
a_{-1,t^{-1}}=a_{-2}a_{-1,x}-a_{-1}a_{-2,x},\text{ \ }%
a_{-2,t^{-1}}=a_{-3}a_{-1,x}-a_{-1}a_{-3,x}
\end{equation*}
together with the first \textquotedblleft negative\textquotedblright\
equation of hydrodynamic chain (\ref{h})%
\begin{equation*}
a_{-1,t^{-2}}=a_{-3}a_{-1,x}-a_{-1}a_{-3,x},
\end{equation*}%
and eliminating the common block $a_{-3}a_{-1,x}-a_{-1}a_{-3,x}$, one can
obtain three dimensional quasilinear system (\ref{odin}).

Instead of asymptotic expansion (\ref{ya}), we can consider an asymptotic
expansion at any fixed point $\zeta =\lambda _{0}$. If we choose $\lambda
_{0}=1$, then instead of (\ref{nega}), we obtain (see (\ref{ful})) $%
a=b_{-1}(\lambda -1)^{-1}$, where $b_{-1}=\mathbf{a}(x,\mathbf{t},1)$; if we
keep $\lambda _{0}$ as an arbitrary constant, then we have another choice $%
a=c_{-1}(\lambda -\lambda _{0})^{-1}$, where $c_{-1}=\mathbf{a}(x,\mathbf{t}%
,\lambda _{0})$. Thus, we must introduce two extra independent variables $%
y^{-1},z^{-1}$ such that two extra generating functions of conservation laws%
\begin{equation*}
p_{y^{-1}}=-\left( \frac{b_{-1}}{\lambda -1}p\right) _{x},\text{ \ }%
p_{z^{-1}}=-\left( \frac{c_{-1}}{\lambda -\lambda _{0}}p\right) _{x}
\end{equation*}%
can be determined as a part of an integrable hierarchy, which contains
hydrodynamic chain (\ref{a}) together with all its commuting flows. This
means that all functions $a_{k}$ (see, for instance, (\ref{a}), (\ref{b}), (%
\ref{ya})), all functions $u^{k}$ (see, for instance, (\ref{mth})) and these
three extra functions $b_{-1},c_{-1}$ and $p$ depend on \textquotedblleft
time\textquotedblright\ variables $t^{k}$ and $y^{-1},z^{-1}$
simultaneously. Consistency of two generating functions of conservation laws%
\begin{equation*}
p_{t^{-1}}=-\frac{1}{\lambda }(a_{-1}p)_{x},\text{ \ }p_{y^{-1}}=-\left(
\frac{b_{-1}}{\lambda -1}p\right) _{x}
\end{equation*}%
yields the fourth three dimensional quasilinear system%
\begin{equation}
a_{-1,y^{-1}}=b_{-1,t^{-1}},\text{ \ }%
a_{-1,y^{-1}}=b_{-1}a_{-1,x}-a_{-1}b_{-1,x}.  \label{four}
\end{equation}

\textbf{Remark}: Introducing a potential function $W$ such that $%
a_{1}=W_{x},a_{2}=W_{t},a_{-1}=W_{t^{-1}},a_{-2}=W_{t^{-2}},b_{-1}=W_{y^{-1}}
$, systems (\ref{three}), (\ref{dva}), (\ref{odin}), (\ref{four}) can be
written as four three dimensional quasilinear equations of a second order
(let us remind that $x=t^{0},t=t^{1},y=t^{2}$), respectively%
\begin{equation}
W_{t^{0}}W_{t^{0}t^{1}}+W_{t^{0}t^{2}}=W_{t^{1}}W_{t^{0}t^{0}}+W_{t^{1}t^{1}},%
\text{ \ }W_{t^{1}t^{-1}}=W_{t^{0}}W_{t^{0}t^{-1}}-W_{t^{-1}}W_{t^{0}t^{0}},
\label{moss}
\end{equation}%
\begin{equation*}
W_{t^{-1}t^{-1}}=W_{t^{-2}}W_{t^{0}t^{-1}}-W_{t^{-1}}W_{t^{0}t^{-2}},\text{
\ }W_{t^{-1}y^{-1}}=W_{y^{-1}}W_{t^{0}t^{-1}}-W_{t^{-1}}W_{t^{0}y^{-1}}.
\end{equation*}

Now we consider three generating functions of conservation laws%
\begin{equation*}
p_{t^{-1}}=-\frac{1}{\lambda }(a_{-1}p)_{x},\text{ \ }p_{y^{-1}}=-\frac{1}{%
\lambda -1}(b_{-1}p)_{x},\text{ \ }p_{z^{-1}}=-\frac{1}{\lambda -\lambda _{0}%
}(c_{-1}p)_{x},
\end{equation*}%
where $a_{-1}=W_{t^{-1}},b_{-1}=W_{y^{-1}}$ and $c_{-1}=W_{z^{-1}}$. One can
introduce a potential function $S$ such that%
\begin{equation*}
dS=pdx-\frac{1}{\lambda }a_{-1}pdt^{-1}-\frac{b_{-1}}{\lambda -1}pdy^{-1}-%
\frac{c_{-1}}{\lambda -\lambda _{0}}pdz^{-1}.
\end{equation*}%
Then the compatibility conditions $%
(S_{t^{-1}})_{y^{-1}}=(S_{y^{-1}})_{t^{-1}}$ and $%
(S_{t^{-1}})_{z^{-1}}=(S_{z^{-1}})_{t^{-1}}$ yield%
\begin{equation*}
q_{y^{-1}}=\left( \frac{\lambda }{\lambda -1}\frac{b_{-1}}{a_{-1}}q\right)
_{t^{-1}},\text{ \ }q_{z^{-1}}=\left( \frac{\lambda }{\lambda -\lambda _{0}}%
\frac{c_{-1}}{a_{-1}}q\right) _{t^{-1}},
\end{equation*}%
where $q=a_{-1}p/\lambda $. The compatibility condition $%
(S_{z^{-1}})_{y^{-1}}=(S_{y^{-1}})_{z^{-1}}$ is equivalent to the
compatibility condition $(q_{z^{-1}})_{y^{-1}}=(q_{y^{-1}})_{z^{-1}}$, which
implies the fifth three dimensional quasilinear system%
\begin{equation}
(\lambda
_{0}-1)W_{t^{-1}}W_{y^{-1}z^{-1}}+W_{z^{-1}}W_{y^{-1}t^{-1}}-\lambda
_{0}W_{y^{-1}}W_{z^{-1}t^{-1}}=0,\text{ \ }  \label{fermos}
\end{equation}

Recently a complete classification of linearly degenerate three dimensional
quasilinear equations of a second order was presented in \cite{fermos}. The
list of these equations\footnote{%
all these three dimensional quasilinear equations of a second order can be
found, for instance, in \cite{maksjmp} and \cite{mas2}. Some other
references are in \cite{fermos}.} coincides with five equations (\ref{moss}%
), (\ref{fermos}).

Since all these five equations belong to the same hierarchy (the function $W$
is common), the computation of differential reductions of last four three
dimensional quasilinear equations of a second order reduces to the problem
of computation of higher (positive) and lower (negative) commuting flows of
integrable dispersive systems (\ref{disp}). For instance, second three
dimensional quasilinear system (\ref{dva}) possesses infinitely many
differential constraints $a_{1}(\mathbf{u},\mathbf{u}_{x},...),a_{-1}(%
\mathbf{u},\mathbf{u}_{x},...)$ (see expressions (\ref{a1}), (\ref{ogra}), (%
\ref{okra})), where field variables $u^{k}(x,t,t^{-1})$ are solutions of
dispersive systems (\ref{disp}) and (\ref{nonloc}). Moreover all lower
differential constraints $a_{-k}(\mathbf{u},\mathbf{u}_{x},...)$ can be
found directly from (\ref{kvadro}) by substitution the expansion (cf. (\ref%
{raz}))%
\begin{equation*}
\varphi (x,\mathbf{t},\lambda )=a_{-1}+\lambda a_{-2}+\lambda ^{2}a_{-3}+...,%
\text{ \ }\lambda \rightarrow 0,
\end{equation*}%
together with (\ref{rac}). Indeed, the ordinary differential equation on the
function $a_{-1}$ coincides with (\ref{ogra}). The next coefficient $a_{-2}$
is determined by the ordinary differential equation\footnote{%
constants $\xi _{0},\xi _{1}$ are first coefficients of the expansion $%
\epsilon ^{2}(\lambda )=\xi _{0}+\lambda \xi _{1}+...$ (see (\ref{kvadro})).
Obviously, the coefficient $\tilde{\xi}$ coincides with $\xi _{0}$.}%
\begin{equation}
(-1)^{M}\frac{\overset{M+K}{\underset{m=1}{\dprod }}s^{m}}{\overset{K}{%
\underset{k=1}{\dprod }}r^{k}}\left( \overset{K}{\underset{p=1}{\sum }}\frac{%
1}{r^{p}}-\overset{M+K}{\underset{p=1}{\sum }}\frac{1}{s^{p}}\right) =\frac{%
a_{-2,xx}}{2a_{-1}}-\frac{a_{-1,x}}{2a_{-1}^{2}}a_{-2,x}  \label{opra}
\end{equation}%
\begin{equation*}
+\left( \frac{a_{-1,x}^{2}}{2a_{-1}^{3}}-\frac{a_{-1,xx}}{2a_{-1}^{2}}-\frac{%
2\xi _{0}}{4a_{-1}^{3}}\right) a_{-2}+\frac{\xi _{1}}{4a_{-1}^{2}},
\end{equation*}%
where the function $a_{-1}$ as already determined by (\ref{ogra}). Thus
third three dimensional quasilinear system (\ref{odin}) possesses
differential reductions (\ref{nonloc}) and%
\begin{equation*}
r_{t^{-2}}^{k}=-\left( \frac{a_{-2}}{r^{k}}+\frac{a_{-1}}{(r^{k})^{2}}%
\right) r_{x}^{k},
\end{equation*}%
\begin{equation*}
s_{t^{-2}}^{i}=-\left( \frac{a_{-2}}{s^{i}}+\frac{a_{-1}}{(s^{i})^{2}}%
\right) s_{x}^{i}-\frac{1}{2(s^{i})^{2}}\frac{\overset{K}{\underset{k=1}{%
\dprod }}(s^{i}-r^{k})}{\underset{m\neq i}{\dprod }(s^{i}-s^{m})}%
(s^{i}a_{-2,xxx}+a_{-1,xxx}),
\end{equation*}%
where differential constraints $a_{-1}(\mathbf{u},\mathbf{u}_{x},...),a_{-2}(%
\mathbf{u},\mathbf{u}_{x},...)$ are determined by (\ref{ogra}) and (\ref%
{opra}). More lengthy computations lead to similar but more complicated
expressions for fourth three dimensional quasilinear system (\ref{four}) and
for the fifth three dimensional quasilinear equation of a second order (\ref%
{fermos}). We omit corresponding derivations here.

\section{The Dispersionless Limit}

\label{sec:dispersionless}

Hydrodynamic chain (\ref{a}) also can be written in the form (see formula
(53) in \cite{maksjmp})%
\begin{equation}
u_{t}^{k}=u_{x}^{k+1}-\frac{1}{2}u^{1}u_{x}^{k}-u^{k}u_{x}^{1}.
\label{hydro}
\end{equation}%
Indeed, the substitution $\varphi =q^{-1/2}$ into the second equation in (%
\ref{k}) yields%
\begin{equation*}
q_{t}=\left( \lambda -\frac{1}{2}u^{1}\right) q_{x}-qu_{x}^{1},
\end{equation*}%
where $a_{1}=-u^{1}/2$ and (taking into account (\ref{raz}))%
\begin{equation*}
q=1+\frac{u^{1}}{\lambda }+\frac{u^{2}}{\lambda ^{2}}+\frac{u^{3}}{\lambda
^{3}}+...
\end{equation*}%
Thus hydrodynamic chain (\ref{hydro}) is a dispersionless limit of $M$th
dispersive chain (\ref{mth}) for any $M>0$ (also this statement is valid for
$M=0$, see Section \ref{sec:exception}). So, we come to an interesting
observation: hydrodynamic chain (\ref{a}) under infinitely many triangular
invertible transformations (\ref{con}) can be written as dispersive chain (%
\ref{mth}) (by infinitely many different ways depended on the natural number
$M$), whose dispersionless limit (\ref{hydro}) is equivalent to original
hydrodynamic chain (\ref{a}) up to invertible triangular point
transformations. Hydrodynamic chain (\ref{hydro}) and its higher commuting
flows\footnote{%
these commuting flows are dispersionless limit of (\ref{com}). Here
coefficients (see (\ref{observ})) $a_{m}=\frac{\partial h_{m+s}}{\partial
u^{s}},m=0,1,...,s=1,2,...$, where Hamiltonian densities $h_{k}(\mathbf{u})$
are determined below.} have infinitely many local Hamiltonian structures.
They can be easily obtain by dispersionless limit from local Hamiltonian
structures of $M$th dispersive chain (see Subsection \ref{subsec:hamilton}).
First three of them are ($s=1,2,...$)%
\begin{equation*}
u_{t^{s}}^{k}=\overset{s+1}{\underset{m=1}{\sum }}(u^{k+m-1}\partial
_{x}+\partial _{x}u^{k+m-1})\frac{\partial h_{s+1}}{\partial u^{m}},\text{ }%
k=1,2,...;
\end{equation*}%
\begin{equation*}
u_{t^{s}}^{1}=-2\partial _{x}\frac{\partial h_{s+2}}{\partial u^{1}},\text{
\ }u_{t^{s}}^{k}=\overset{s+2}{\underset{m=2}{\sum }}(u^{k+m-2}\partial
_{x}+\partial _{x}u^{k+m-2})\frac{\partial h_{s+2}}{\partial u^{m}},\text{ }%
k=2,3,...;
\end{equation*}

\begin{equation*}
u_{t^{s}}^{1}=-2\partial _{x}\frac{\partial h_{s+3}}{\partial u^{2}},\text{
\ }u_{t^{s}}^{2}=-2\partial _{x}\frac{\partial h_{s+3}}{\partial u^{1}}%
-(u^{1}\partial _{x}+\partial _{x}u^{1})\frac{\partial h_{s+3}}{\partial
u^{2}},
\end{equation*}%
\begin{equation*}
u_{t^{s}}^{k}=\overset{s+3}{\underset{m=3}{\sum }}(u^{k+m-3}\partial
_{x}+\partial _{x}u^{k+m-3})\frac{\partial h_{s+3}}{\partial u^{m}},\text{ }%
k=3,4,...
\end{equation*}%
All Hamiltonian densities $h_{k}(\mathbf{u})$ can be found from (\ref{pi})
by substitution $p=1/\varphi $, where expansion of $\varphi $ is determined
by (\ref{raz}). For instance first conservation law densities are
(corresponding conservation laws follow from (\ref{pi}) and (\ref{hydro}),
see other detail in \cite{maksjmp})%
\begin{equation*}
h_{1}=u^{1},\text{ \ }h_{2}=u^{2}-\frac{1}{4}(u^{1})^{2},\text{ \ }%
h_{3}=u^{3}-\frac{1}{2}u^{1}u^{2}+\frac{1}{8}(u^{1})^{3},
\end{equation*}

\begin{equation*}
h_{4}=u^{4}-\frac{1}{2}u^{1}u^{3}-\frac{1}{4}(u^{2})^{2}+\frac{3}{8}%
(u^{1})^{2}u^{2}-\frac{5}{64}(u^{1})^{4},...
\end{equation*}

A dispersionless limit of dispersive reductions associated with ansatz (\ref%
{racio}) was investigated in \cite{coupledKdV}. In this Section we consider
a dispersionless limit of dispersive reductions (\ref{disp}), i.e. the
dispersionless systems%
\begin{equation*}
r_{t}^{k}=(r^{k}+a_{1})r_{x}^{k},\text{ \ }s_{t}^{i}=(s^{i}+a_{1})s_{x}^{i},
\end{equation*}%
where%
\begin{equation*}
a_{1}=\frac{1}{2}\left( \overset{M+K}{\underset{m=1}{\sum }}s^{m}-\overset{K}%
{\underset{k=1}{\sum }}r^{k}\right) .
\end{equation*}%
Thus the field variables $r^{k},s^{m}$ are Riemann invariants (cf. formula
(6) in \cite{coupledKdV}). These coordinates $r^{k},s^{m}$ are most
convenient for comparison of distinct integrable systems. For instance,
Kaup--Boussinesq system (\ref{kb}), Ito system (\ref{ito}) and two component
generalization of the Camassa--Holm equation (\ref{ch2}) have the same
dispersionless limit (see the end of Subsection \ref{subsec:racio})%
\begin{equation*}
s_{t}^{1}=\frac{1}{2}(3s^{1}+s^{2})s_{x}^{1},\text{ \ \ \ }s_{t}^{2}=\frac{1%
}{2}(s^{1}+3s^{2})s_{x}^{2}.
\end{equation*}%
However, these three integrable dispersive systems do not connected by any
differential substitutions. They are different, but have the same
dispersionless limit.

In paper \cite{maksjmp} we proved that hydrodynamic chain (\ref{a})
possesses $N$ component hydrodynamic reductions%
\begin{equation}
r_{t}^{i}=(r^{i}+a_{1})r_{x}^{i},  \label{rim}
\end{equation}%
where ($f_{k}(r^{k})$ are arbitrary functions)%
\begin{equation*}
a_{1}=\overset{N}{\underset{m=1}{\sum }}f_{m}(r^{m}).
\end{equation*}%
Thus, in this paper we proved that hydrodynamic type system (\ref{rim}) has
a dispersive integrable extension (\ref{disp}) if%
\begin{equation*}
a_{1}=\frac{1}{2}\overset{N}{\underset{m=1}{\sum }}\epsilon _{m}r^{m},
\end{equation*}%
where $\epsilon _{k}=\pm 1$, while in \cite{coupledKdV} all constants $%
\epsilon _{k}=1$.

\textbf{Remark}\footnote{%
this derivation of the relationship between the Kaup--Boussinesq system (\ref%
{kb}), the Kaup--Broer system (\ref{kb2}) and the nonlinear Schr\"{o}dinger
equation (\ref{kb3}) belongs to Solomon J. Alber.}: Kaup--Boussinesq system (%
\ref{kb}) can be written in the conservative form%
\begin{equation}
u_{t}^{1}=\left( w-\frac{1}{2}(u^{1})^{2}\right) _{x},\text{ \ }w_{t}=\left(
\frac{1}{4}u_{xx}^{1}-u^{1}w\right) _{x},  \label{kb1}
\end{equation}%
where%
\begin{equation*}
w=u^{2}-\frac{1}{4}(u^{1})^{2}.
\end{equation*}%
Under the invertible substitution%
\begin{equation*}
w=\rho \pm \frac{1}{2}u_{x}^{1},
\end{equation*}%
(\ref{kb1}) becomes the Kaup--Broer system%
\begin{equation}
u_{t}^{1}=\left( \rho -\frac{1}{2}(u^{1})^{2}\pm \frac{1}{2}u_{x}^{1}\right)
_{x},\text{ \ }\rho _{t}=-\left( u^{1}\rho \pm \frac{1}{2}\rho _{x}\right)
_{x}.  \label{kb2}
\end{equation}%
Under the invertible substitution%
\begin{equation*}
v=u^{1}\pm \frac{1}{2}\frac{\rho _{x}}{\rho },
\end{equation*}%
(\ref{kb2}) reduces to the Hasimoto form of the nonlinear Schr\"{o}dinger
equation%
\begin{equation}
v_{t}=\left( \rho -\frac{v^{2}}{2}-\frac{\rho _{xx}}{4\rho }+\frac{\rho
_{x}^{2}}{8\rho ^{2}}\right) _{x},\text{ \ }\rho _{t}=-(v\rho )_{x}.
\label{kb3}
\end{equation}%
Indeed, under the Madelung substitution%
\begin{equation*}
\psi =\sqrt{\rho }e^{\int vdx},\text{ \ \ }\psi ^{+}=\sqrt{\rho }e^{-\int
vdx},
\end{equation*}%
(\ref{kb3}) transforms into the nonlinear Schr\"{o}dinger equation%
\begin{equation}
\psi _{t}+\frac{1}{2}\psi _{xx}-\psi \psi ^{+}\psi =0,\text{ \ \ }\psi
_{t}^{+}-\frac{1}{2}\psi _{xx}^{+}+\psi ^{+}\psi \psi ^{+}=0.  \label{kb4}
\end{equation}%
Thus nonlinear Schr\"{o}dinger equation (\ref{kb4}) is reducible to the form
(\ref{j}), where%
\begin{equation*}
\rho =-\frac{1}{4}(s^{1}-s^{2})^{2}\pm \frac{1}{2}(s^{1}+s^{2})_{x},\text{ \
}v=-s^{1}-s^{2}\pm \frac{1}{2}(\ln \rho )_{x}.
\end{equation*}%
Also nonlinear Schr\"{o}dinger equation (\ref{kb4}) is well known in the
complex form%
\begin{equation}
i\psi _{t}-\frac{1}{2}\psi _{xx}-\psi \psi ^{+}\psi =0,\text{ \ \ }i\psi
_{t}^{+}+\frac{1}{2}\psi _{xx}^{+}+\psi ^{+}\psi \psi ^{+}=0,  \label{kb5}
\end{equation}%
which follows from (\ref{kb4}) by the complex change of independent
variables $x\rightarrow ix,t\rightarrow it$. In such a case (\ref{kb5}) can
be written in the form (cf. (\ref{j}))%
\begin{equation*}
s_{t}^{1}=\frac{1}{2}(3s^{1}+s^{2})s_{x}^{1}-\frac{(s^{1}+s^{2})_{xxx}}{%
4(s^{1}-s^{2})},\text{ \ \ }s_{t}^{2}=\frac{1}{2}(s^{1}+3s^{2})s_{x}^{2}+%
\frac{(s^{1}+s^{2})_{xxx}}{4(s^{1}-s^{2})}.
\end{equation*}%
Thus, all above systems possess the same dispersionless limit. The field
variables $s^{1},s^{2}$ are natural coordinates for the nonlinear Schr\"{o}%
dinger equation. For investigation of shock waves ($\epsilon \rightarrow 0$)
the nonlinear Schr\"{o}dinger equation can be written in the form (by
rescaling $\partial _{x}\rightarrow \epsilon \partial _{x},\partial
_{t}\rightarrow \epsilon \partial _{t}$)%
\begin{equation*}
s_{t}^{1}=\frac{1}{2}(3s^{1}+s^{2})s_{x}^{1}-\epsilon ^{2}\frac{%
(s^{1}+s^{2})_{xxx}}{4(s^{1}-s^{2})},\text{ \ \ }s_{t}^{2}=\frac{1}{2}%
(s^{1}+3s^{2})s_{x}^{2}+\epsilon ^{2}\frac{(s^{1}+s^{2})_{xxx}}{%
4(s^{1}-s^{2})}.
\end{equation*}%
Such a consideration obviously is valid just if the difference $s^{1}-s^{2}$
is not small.

\section{Conclusion}

\label{sec:conclusion}

In this paper we found new multi component integrable dispersive systems (%
\ref{disp}) associated with the energy dependent Schr\"{o}dinger equation.
We believe that the same approach based on construction of dispersive chains
can be utilized for all others linear spectral problems. Moreover we hope
that each linear spectral problem can be associated with some three
dimensional linearly degenerate quasilinear system of a first order. Thus we
would like to establish a link between three dimensional linearly degenerate
quasilinear systems of a first order and two dimensional dispersive
integrable systems.

More general alternative approach for extraction of integrable dispersive
systems from (\ref{main}) is based on extension of method of hydrodynamic
reductions (see detail in \cite{FerMoro}) to integrable dispersive
deformations. In this case functions $a_{1}$ and $a_{2}$ (see (\ref{three}))
depend on $N$ Riemann invariants $r^{k}$ and can be written in the
quasipolynomial form%
\begin{equation*}
a_{1}=a_{10}(\mathbf{r})+\epsilon b_{1k}(\mathbf{r})r_{x}^{k}+\epsilon
^{2}(c_{1km}(\mathbf{r})r_{x}^{k}r_{x}^{m}+c_{1k}(\mathbf{r})r_{xx}^{k})+...,
\end{equation*}%
\begin{equation*}
a_{2}=a_{20}(\mathbf{r})+\epsilon b_{2k}(\mathbf{r})r_{x}^{k}+\epsilon
^{2}(c_{2km}(\mathbf{r})r_{x}^{k}r_{x}^{m}+c_{2k}(\mathbf{r})r_{xx}^{k})+...,
\end{equation*}%
where Riemann invariants $r^{k}(x,t,y)$ satisfy two commuting systems%
\begin{equation}
r_{t}^{i}=\lambda ^{i}(\mathbf{r})r_{x}^{i}+\epsilon (\lambda _{km}^{i}(%
\mathbf{r})r_{x}^{k}r_{x}^{m}+\lambda _{k}^{i}(\mathbf{r})r_{xx}^{k})+...
\label{r1}
\end{equation}%
\begin{equation}
r_{y}^{i}=\mu ^{i}(\mathbf{r})r_{x}^{i}+\epsilon (\mu _{km}^{i}(\mathbf{r}%
)r_{x}^{k}r_{x}^{m}+\mu _{k}^{i}(\mathbf{r})r_{xx}^{k})+...  \label{r2}
\end{equation}%
Functions $a_{10}(\mathbf{r}),b_{1k}(\mathbf{r}),\lambda ^{i}(\mathbf{r}%
),... $ are not yet determined. However, substitution of these expressions
into (\ref{three}) leads to (see detail in \cite{maksjmp})%
\begin{equation*}
a_{10}(\mathbf{r})=\overset{N}{\underset{m=1}{\sum }}\Psi _{m}(r^{m}),\text{
\ }a_{20}(\mathbf{r})=\frac{1}{2}a_{10}^{2}+\overset{N}{\underset{m=1}{\sum }%
}\int f_{m}(r^{m})\Psi _{m}(r^{m})dr^{m},
\end{equation*}%
\begin{equation*}
\lambda ^{i}(\mathbf{r})=f_{i}(r^{i})+a_{10},\text{ \ }\mu ^{i}(\mathbf{r}%
)=f_{i}^{2}(r^{i})+a_{10}f_{i}(r^{i})+a_{20}.
\end{equation*}%
All other coefficients $c_{1km}(\mathbf{r}),c_{1k}(\mathbf{r}),\lambda
_{km}^{i}(\mathbf{r}),\lambda _{k}^{i}(\mathbf{r}),...$ can be found
iteratively for each degree of the parameter $\epsilon $. Integrability of
systems (\ref{r1}), (\ref{r2}) follows from (\ref{main}). Corresponding
equations%
\begin{equation*}
u_{t}=2ua_{1,x}+(\lambda +a_{1})u_{x}-\frac{1}{2}\epsilon ^{2}a_{1,xxx},
\end{equation*}%
\begin{equation*}
u_{y}=2u(\lambda a_{1,x}+a_{2,x})+(\lambda ^{2}+\lambda a_{1}+a_{2})u_{x}-%
\frac{1}{2}\epsilon ^{2}(\lambda a_{1,xxx}+a_{2,xxx})
\end{equation*}%
determine a common function $u(x,t,y,\lambda )$, which one can look for in
the form%
\begin{equation*}
u=u_{0}(\mathbf{r},\lambda )+\epsilon u_{1k}(\mathbf{r},\lambda
)r_{x}^{k}+\epsilon ^{2}[u_{2k}(\mathbf{r},\lambda )r_{xx}^{k}+u_{2km}(%
\mathbf{r},\lambda )r_{x}^{k}r_{x}^{m}]+...
\end{equation*}%
This investigation should be made in a separate paper.

\section*{Acknowledgement}

MVP's work was also partially supported by the RF Government grant
\#2010-220-01-077, ag. \#11.G34.31.0005, by the grant of Presidium of RAS
\textquotedblleft Fundamental Problems of Nonlinear
Dynamics\textquotedblright\ and by the RFBR grant 14-01-00389. MVP thanks
E.V. Ferapontov and G.A. El for useful comments and important discussions.


\addcontentsline{toc}{section}{References}

\begin{thebibliography}{99}
\bibitem{AdlerShabat} \emph{V. E. Adler, A.B. Shabat,} \newblock Model
equation of the theory of solitons, \textit{Theor. Math. Phys.} \textbf{153}
No. 1 (2007) 1373--1387.

\bibitem{SAlber} \emph{S.J. Alber,} \newblock Hamiltonian systems on the
Jacobi varieties. The geometry of Hamiltonian systems (Berkeley, CA, 1989)
23--32, Math. Sci. Res. Inst. Publ., 22, Springer, New York, 1991. \emph{%
S.J. Alber,} \newblock Associated integrable systems, \textit{J. Math. Phys}%
. \textbf{32} No. 4 (1991) 916-922.

\bibitem{antford} \emph{M. Antonowicz, A. Fordy,} \newblock A family of
completely integrable multi-Hamiltonian systems, \textit{Phys. Letts. A}
\textbf{122} (1987) 95--99. \emph{M. Antonowicz, A. Fordy,} \newblock %
Coupled KdV equations with multi-Hamiltonian structures, \textit{Physica}
\textbf{28D} (1987) 345--57.

\bibitem{antford2} \emph{M. Antonowicz, A. Fordy,} \newblock Coupled Harry
Dym equations with multi-Hamiltonian structures, \textit{J. Phys. A} \textbf{%
21} (1988) L269--75. \emph{M. Antonowicz, A. Fordy,} \newblock Factorisation
of energy dependent Schr\"{o}dinger operators: Miura maps and modified
systems, \textit{Comm. Math. Phys}., \textbf{124} (1989) 465-486.

\bibitem{Fer91} \emph{E.V. Ferapontov}, \newblock Integration of
weakly-nonlinear hydrodynamic systems in Riemann invariants, \textit{Phys.
Lett. A} \textbf{158} (1991) 112-118.

\bibitem{ferap} \emph{E.V. Ferapontov, K.R. Khusnutdinova, D.G. Marshall,
M.V. Pavlov}, \newblock Classification of Integrable Hydrodynamic chains
associated with Kupershmidt's brackets, \textit{J. Math. Phys}., \textbf{47}
(2006) 103507-103520. \emph{E.V. Ferapontov, D.G. Marshall}, \newblock %
Differential-geometric approach to the integrability of hydrodynamic chains:
the Haantjes tensor, \textit{Mathematische Annalen}, \textbf{339} No. 1
(2007) 61-99.

\bibitem{FerMoro} \emph{E.V. Ferapontov, A. Moro}, \newblock Dispersive
deformations of hydrodynamic reductions of (2 + 1)D dispersionless
integrable systems, \textit{J. Phys. A} \textbf{42} No. 3 (2009) 035211.
\emph{E.V. Ferapontov, A. Moro, V.S. Novikov}, \newblock Integrable
equations in 2 + 1 dimensions: deformations of dispersionless limits,
\textit{J. Phys. A} \textbf{42} No. 34 (2009) 345205.

\bibitem{fermos} \emph{E.V. Ferapontov, J. Moss}, \newblock Linearly
degenerate PDEs and quadratic line complexes, arXiv:1204.2777, submitted.

\bibitem{coupledKdV} \emph{E.V. Ferapontov, M.V. Pavlov}, \newblock %
Quasiclassical limit of coupled KdV equations. Riemann invariants and
multi-Hamiltonian structure, \textit{Physica }\textbf{52D} No. 2--3 (1991)
211--219.

\bibitem{gibtsar} \emph{J. Gibbons, S.P. Tsarev}, \newblock Reductions of
Benney's equations, \textit{Phys. Lett. A}, \textbf{211 }(1996) 19-24. \emph{%
J. Gibbons, S.P. Tsarev}, \newblock Conformal maps and reductions of the
Benney equations, \textit{Phys. Lett. A}, \textbf{258} (1999) 263-270.

\bibitem{lma} \emph{L. Martinez Alonso,} \newblock Schr\"{o}dinger spectral
problems with energy--dependent potentials as sources of nonlinear
Hamiltonian evolution equations, \textit{J. Math. Phys.} \textbf{21} (1980)
2342--2349.

\bibitem{mas} \emph{L. Martinez Alonso, A.B. Shabat,} \newblock %
Energy-dependent potentials revisited: a universal hierarchy of hydrodynamic
type, \textit{Phys. Letts. A} \textbf{299} No. 4 (2002) 359--365. \emph{L.
Martinez Alonso, A.B. Shabat,} \newblock Towards a Theory of Differential
Constraints of a Hydrodynamic Hierarchy, \textit{J. Nonlin. Math. Phys.}
\textbf{10} No. 2 (2003) 229--242.

\bibitem{mas2} \emph{L. Martinez Alonso, A.B. Shabat,} \newblock %
Hydrodynamic Reductions and Solutions of a Universal Hierarchy, \textit{%
Theor. Math. Phys.} \textbf{140 }No. 2 (2004) 1073--1085.

\bibitem{marvan} \emph{M. Marvan, M.V. Pavlov,} \newblock Energy dependent
Schr\"{o}dinger operator and a new subclass of integrable systems, submitted
to SIGMA.

\bibitem{maksjmp} \emph{M.V. Pavlov,} \newblock Integrable hydrodynamic
chains, \textit{J. Math. Phys}. \textbf{44} No. 9 (2003) 4134-4156.

\bibitem{Makslin} \emph{M.V. Pavlov,} \newblock Hamiltonian formalism of
weakly nonlinear systems in hydrodynamics, \textit{Theor. Math. Phys.}
\textbf{73} (1987) 1242--1245.

\bibitem{makschain} \emph{M.V. Pavlov}, \newblock The Kupershmidt
hydrodynamic chains and lattices, \textit{IMRN} (2006) article ID 46987.
\emph{M.V. Pavlov}, \newblock Classification of integrable hydrodynamic
chains and generating functions of conservation laws, \textit{J. Phys. A:
Math. Gen}., (2006) 10803-10819. \emph{M.V. Pavlov}, \newblock %
Algebro-geometric approach in the theory of integrable hydrodynamic type
systems, \textit{Comm. Math. Phys}., \textbf{272} No. 2 (2007) 469-505.
\emph{M.V. Pavlov}, \newblock The Hamiltonian approach in the classification
and the integrability of hydrodynamic chains, ArXiv: Nlin.SI/0603057. \emph{%
M.V. Pavlov, S.A. Zykov}, \newblock Classification of integrable
conservative hydrodynamic chains, submitted (2010) arXiv: 0912.4954.

\bibitem{ShabatJNMP} \emph{A.B. Shabat,} \newblock Universal solitonic
hierarchy, \textit{J. Nonlin. Math. Phys.} \textbf{12} Supplement 1 (2005)
614--624. \emph{A.B. Shabat,} \newblock Symmetric polynomials and
conservation laws, \textit{Vladikavkazskij matematicheskij zhurnal} \textbf{%
14} No. 4 (2012) 83--94 (in Russian).
\end{thebibliography}
\end{document}